\documentclass[useAMS,usenatbib]{mn2e}
\usepackage{amsmath}
\usepackage{amssymb}
\usepackage{graphicx}
\usepackage{subfig}
\usepackage{epsfig}
\usepackage{caption}
\usepackage{breqn}
\usepackage{changebar}
\usepackage{xcolor}

\title[The Galaxy Luminosity and Star-Formation Rate Function at High Redshift]{An Empirical Model for the Galaxy Luminosity and Star-Formation Rate Function at High Redshift}
\author[N. Mashian et al.]{Natalie Mashian$^{1}$\thanks{nmashian@physics.harvard.edu},Pascal A. Oesch$^{2}$\thanks{pascal.oesch@yale.edu}, Abraham Loeb$^{1}$\thanks{aloeb@cfa.harvard.edu} \\
$^{1}$Harvard-Smithsonian Center for Astrophysics, 60 Garden Street, Cambridge, MA 02138, USA\\
$^{2}$Yale Center for Astronomy and Astrophysics, Physics and Astronomy Departments, New Haven, CT 06520, USA}

\begin{document}

\pagerange{\pageref{firstpage}--\pageref{lastpage}} \pubyear{2015}

\maketitle

\label{firstpage}


\begin{abstract}
Using the most recent measurements of the ultraviolet (UV) luminosity functions (LFs) and dust estimates of early galaxies, we derive updated dust-corrected star-formation rate functions (SFRFs) at $z\sim4-8$, which we model to predict the evolution to higher redshifts, $z>8$.
We employ abundance matching techniques to calibrate a relation between galaxy star formation rate (SFR) and host halo mass $M_h$ by mapping the shape of the observed SFRFs at $z \sim$ 4-8 to that of the halo mass function. The resulting scaling law remains roughly constant over this redshift range. We apply the average $SFR-M_h$ relation to reproduce the observed SFR functions at 4 $\lesssim z \lesssim$ 8 and also derive the expected UV LFs at higher redshifts. At $z\sim9$ and $z\sim10$ these model LFs are in excellent agreement with current observed estimates.  Our predicted number densities and UV LFs at $z>10$ indicate that JWST will be able to detect galaxies out to $z\sim15$ with an extensive treasury sized program. 
We also derive the redshift evolution of the star formation rate density and associated reionization history by galaxies. Models which integrate down to the current HUDF12/XDF detection limit ($M_{UV} \sim$ -17.7 mag) result in a SFRD that declines as $(1+z)^{-10.4\pm0.3}$ at high redshift and fail to reproduce the observed CMB electron scattering optical depth, $\tau \simeq$ 0.066, to within 1$\sigma$. On the other hand, we find that the inclusion of galaxies with SFRs well below the 
current detection limit ($M_{UV} <$ -5.7 mag)  leads to a fully reionized universe by $z \sim$ 6.5 and an optical depth of $\tau \simeq$ 0.054, consistent with the recently derived Planck value at the 1$\sigma$ level. 
\end{abstract}


\begin{keywords}
cosmology: theory -- galaxies: high-redshift -- galaxies: mass function.
\end{keywords}

\section{Introduction}
\indent The sensitive, near-infrared imaging capabilities of the Hubble Space Telescope (HST) have significantly expanded our understanding of galaxy evolution through cosmic time. Efforts to identify galaxies through photometric selections and follow-up spectroscopy have progressively extended the observational frontier to higher and higher redshifts, resulting in robust galaxy samples in the range $z \sim$ 4-8 \citep[e.g.][]{1996MNRAS.283.1388M,1999ApJ...519....1S,2004ApJ...616L..79B,2010ApJ...709L.133B,2011ApJ...737...90Ba,2014ApJ...793..115Ba,2004ApJ...600L..99D,2004ApJ...612L..93Y,2010MNRAS.409..855B,2010ApJ...719.1250F,2012ApJ...756..164Fa,2012ApJ...758...93Fb,2014arXiv1410.5439F,2010MNRAS.403..960M,2013MNRAS.432.2696M,2010ApJ...709L..21O,2012ApJ...759..135Oa,2012ApJ...744..179S,2013ApJ...768..196S,2014ApJ...792...76B,2014ApJ...786...57S}. With the Wide Field Camera 3 (WFC3/IR) on board the HST, the current frontier for identifying high-redshift galaxies now lies at $z \sim$ 11, a mere half a billion years after the Big Bang, and new surveys are now building up the sample sizes of $z \gtrsim$ 9 galaxies \citep[e.g.][]{2011Natur.469..504Bb,2013ApJ...763L...7E,2012Natur.489..406Z,2013ApJ...762...32C,2013MNRAS.432.2696M,2013ApJ...773...75O,2014ApJ...786..108O,2014ApJ...795..126Bb,2012ApJ...745..110Ob,2015MNRAS.450.3032M,Ishigaki15}.

However, due to the small number statistics, our understanding of galaxies beyond $z \sim$ 8 is still very limited and obtaining accurate constraints on the evolution of  the galaxy rest-frame ultraviolet luminosity function (hereafter referred to as the luminosity function, or UV LF) in the early universe remains challenging. Typically parameterized by a Schechter function with a power-law slope at faint luminosities and an exponentially declining form at the bright end, the luminosity function provides a measure of the relative space density of galaxies over a wide range of luminosities at a particular redshift \citep{1976ApJ...203..297S}. Since the UV 1500 \AA\, light probes recent star formation activity, the integral of the dust-corrected (intrinsic) luminosity function can then be used to derive the cosmic star formation rate density (SFRD) and estimate the galaxy population's contribution to cosmic reionization, a process that appears to have been completed by $z \sim$ 6 \citep{2012ApJ...756..164Fa,loebbook, 2013ApJ...768...71R,Bouwens15}. Given the scarce number of $z \sim$ 9-10 galaxy candidate detections, a reliable determination of the SFRD at these redshifts remains difficult, with conclusions from separate analyses disagreeing on the SFRD evolutionary trends \citep{2013ApJ...762...32C,2013ApJ...763L...7E,2013ApJ...773...75O,2014ApJ...786..108O}. Furthermore, with the identification of galaxies at these redshifts, we are approaching the limit of HST's detection capabilities. Sources beyond $z \sim$ 11 will continue to remain largely inaccessible until the advent of the James Webb Space Telescope (JWST), whose scheduled launch in 2018 promises to yield revolutionary insights into early galaxy formation at $z>10$ \citep{2006SSRv..123..485G}.

The analysis and extrapolation of the UV luminosity function evolution towards higher ($z \approx$ 8) redshifts has been the focus of several past studies, in the form of both hydrodynamical simulations \citep[e.g.][]{2010PASJ...62.1455N,fin,2011MNRAS.414..847S,2012MNRAS.420.1606J,2013MNRAS.434.1486D,Oshea15} and semi-analytical models \citep[e.g.][]{2010ApJ...714L.202T,2012JCAP...04..015M,2013ApJ...768L..37T,2014ApJ...785...65C,2014MNRAS.445.2545D,2015ApJ...799...32B}. 
Successfully reproducing the statistical properties of the observed Lyman-break galaxy (LBG) population at $z$ = 6-9, simulations predict a large number of undetected, low-mass galaxies that may have significantly contributed to the  reionization of the Universe at $z \geq$ 6. Semi-empirical models which tie the evolution of galaxy luminosity to the dark matter properties of the host halos reach similar conclusions, predicting a steepening faint-end slope of the LF with redshift and a sharp decline in the cosmic star formation rate above $z$ =8 \citep{2015arXiv150702685T, 2015arXiv150801204M}.

In this paper, we take advantage of the recent progress in the observed galaxy statistics at redshifts $z \sim$ 4-8 to evolve the luminosity function, along with its derived properties, towards higher redshifts by establishing a link between the galaxy star formation rate (SFR) and its host halo mass via abundance matching techniques \citep{2004MNRAS.353..189V,2006MNRAS.371.1173V, 2009ApJ...696..620C}. Since galaxy formation is governed by the inflow of baryonic matter into the gravitational potential wells of dark matter halos where it cools and initiates star formation, the properties of galaxies are invariably linked to the characteristics of their host halos. We thus seek to derive a scaling relation between dark matter halo mass, $M_h$ and galaxy SFR at each redshift by assuming that there is a one-to-one, monotonic correspondence between these two properties. We anchor our model to the observed luminosity functions at $z \sim$ 4-8 corrected for dust-extinction, mapping their shape to that of the halo mass function at the respective redshifts. Finding that the SFR-$M_h$ scaling law remains roughly constant across this redshift range, we apply the average relation, $SFR_{av}(M_h)$, to reproduce the observed $z \sim$ 9-10 luminosity functions and then further explore how the UV LF, and the corresponding SFRD, evolve at higher redshifts, $z \sim$ 11-20.

We describe the details of our approach and the theoretical framework used to derive the shape and amplitude of the galaxy UV LF in \S2. 
Our predictions for the evolving LF, the resulting SFRD, and the contribution of these galaxy populations to cosmic reionization at high redshifts are presented in \S3.
We conclude in \S4 with a summary of our findings and their implications for future surveys with JWST.  
We adopt a flat, $\Lambda$CDM cosmology with $\Omega_m$ = 0.3, $\Omega_\Lambda$ = 0.7, $\Omega_b$ = 0.045, $H_0$ = 70 km s$^{-1}$Mpc$^{-1}$, i.e. $h$ = 0.7, $\sigma_8$ = 0.82, and $n_s$ = 0.95, consistent with the most recent measurements from Planck \citep{2015arXiv150201589P}.
All magnitudes in this work are in the AB scale \citep{1983ApJ...266..713O}.

\section{The Formalism}

The goal of this paper is to derive an empirical prediction of the evolution of UV LFs at $z>8$ tied to the evolution of dark matter haloes. We do so by first calibrating a $SFR-M_h$ relation at $z\sim4-8$, which we then evolve to $z>8$. Our formalism is described in the following sections.

\subsection{The Observed Star-Formation Rate Functions}
One important limitation of the UV LFs in probing the galaxy build-up in the early universe is dust extinction, which significantly affects the UV luminosities of galaxies. Thanks to deep multi-wavelength photometry with HST, it has become clear that dust extinction evolves with redshift, becoming significantly less important at earlier cosmic times \citep{2009ApJ...705..936B,2012A&A...540A..39C,2012ApJ...756..164Fa,2013MNRAS.432.3520D,2014ApJ...793..115Ba,2014MNRAS.440.3714R}. Thus, the UV LF at high redshift evolves due to two effects: (\textit{i}) evolution in dust extinction, and (\textit{ii}) SFR build-up of the underlying galaxy population. Any model of the UV LFs therefore has to disentangle these two effects. 

Alternatively, we can directly make use of the SFR functions (SFRF), $\phi(SFR, z)$. These provide the number density of galaxies with a given SFR, and are thus corrected for dust extinction. Following \citet{2012ApJ...756...14S}, we derive the observed SFR functions based on the most recent UV LFs at $z>4$ taken from \citet{Bouwens15}, which we correct for dust extinction using the observed UV continuum slopes $\beta$ and their relation with UV extinction \citep{Meurer99}. The UV continuum slopes are modeled as a function of luminosity using the most recent relations from \citet{2014ApJ...793..115Ba}.  For more details on this approach see \citet{2012ApJ...756...14S}.

The dust correction effectively shifts the LF to higher luminosities and slightly lowers the volume densities due to the renormalization of the magnitude bins. The dust-corrected, intrinsic UV luminosities are then converted to SFRs using the following empirical relation \citep{1998ARA&A..36..189K},

\begin{equation}
\frac{\text{SFR}}{M_\odot\,\text{yr$^{-1}$}} = 1.25\times10^{-28} \frac{L_{\text{UV,corr}}}{\text{erg s$^{-1}$Hz$^{-1}$}}
\end{equation}
and the resulting SFR functions are represented with the typical Schechter parameterization,
\begin{equation}
\phi(\text{SFR})\,d\,\text{SFR}=\phi^*\left(\frac{\text{SFR}}{\text{SFR$^*$}}\right)^\alpha \exp{\left(-\frac{\text{SFR}}{\text{SFR$^*$}}\right)}\frac{d\,\text{SFR}}{\text{SFR$^*$}} .
\end{equation}

\begin{table}
\centering
\begin{minipage}{140mm}
  \caption{Schechter Parameters Determined for the SFR Functions \label{tab:sfrf}}
   \begin{tabular}{lccc}
  \hline
  \hline
  $< z >$ & $\log_{10}$$\frac{\phi^*}{\text{Mpc$^{-3}$}}$ & $\log_{10}$$\frac{SFR^*}{M_\odot\text{yr$^{-1}$}}$ & $\alpha$\\
 \hline
 3.8 & -2.79$\pm$0.07 & 1.61$\pm$0.06 & -1.53$\pm$0.03\\
 4.9 & -3.25$\pm$0.10 & 1.75$\pm$0.09 & -1.59$\pm$0.08\\
 5.9 & -3.45$\pm$0.16 & 1.62$\pm$0.14 & -1.62$\pm$0.08\\
 6.8 & -3.67$\pm$0.23 & 1.54$\pm$0.20 & -1.76$\pm$0.12\\
 7.9 & -3.79$\pm$0.31 & 1.31$\pm$0.36 & -1.79$\pm$0.18\\
\hline
\end{tabular}
\end{minipage}
\end{table}

\begin{figure}
\includegraphics[width=84mm]{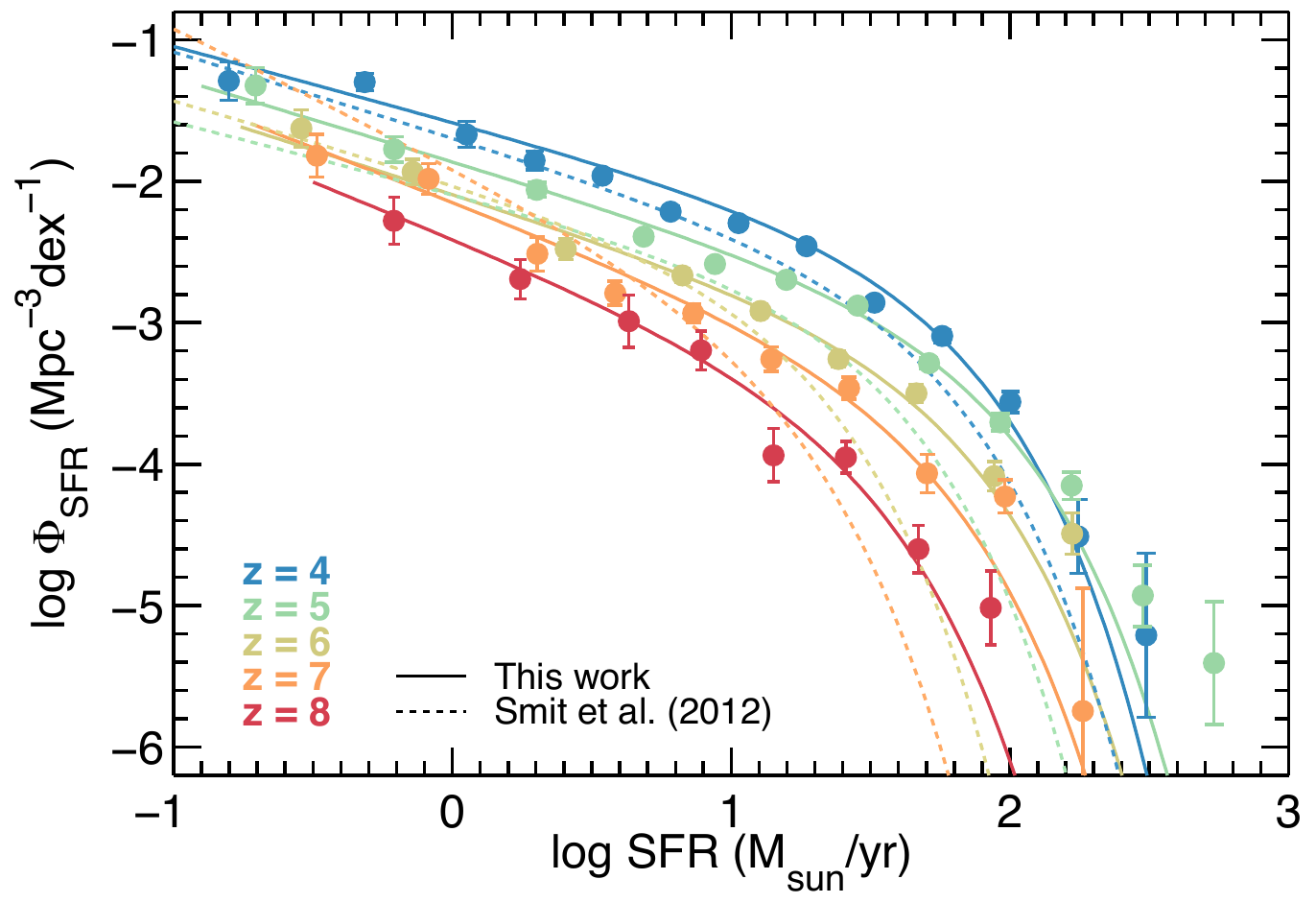}
\caption{The evolution of the star-formation rate function, $\Phi_{\mathrm{SFR}}$, across redshift $z=4-8$. The points correspond to the dust-corrected stepwise UV LFs from \citet{Bouwens15}, which were corrected for dust extinction based on the UV continuum slope distributions measured in \citet{2014ApJ...793..115Ba}. The solid lines show the corresponding Schechter parameters for the SFRFs. These can be compared to the dashed lines from \citet{2012ApJ...756...14S}, which were derived using the same procedure, but based on now-outdated UV LFs and $\beta$ distributions (covering only $z=4-7$). The combination of less evolution in the characteristic luminosity found in newer UV LF results as well as the consistently redder UV continuum slopes, result in significantly higher SFRFs compared to the previous analysis, in particular at higher redshift.}
\label{fig:sfrfComparison}
\end{figure}

The Schechter parameters for the $z \sim$ 4-8 SFR functions are directly related to the Schechter function parameters of the UV LF and the slope of the  $L_{UV}-\beta$ relation \citep[see sections 2.1 and 2.2 in][]{2012ApJ...756...14S}. 
Using the most recent UV LFs and UV continuum slope measurements, we thus update the previous SFR functions of \citet{2012ApJ...756...14S}. The resulting Schechter function parameters are listed in Table \ref{tab:sfrf}. They are also shown in Figure \ref{fig:sfrfComparison}, where we compare our results with the previous analysis of \citet{2012ApJ...756...14S}. As can be seen, the 
combination of less evolution in the characteristic luminosity found in newer UV LF results, as well as the consistently redder UV continuum slopes found in updated measurements, result in significantly higher SFRFs compared to the previous analysis, in particular at higher redshift.

We perform the same analysis also for $z\sim9-10$ galaxy SFR functions. These are much more uncertain, given the very large uncertainties in the UV LFs at these redshifts. Also, note that based on the evolution of the UV continuum slope distribution at lower redshift \citep{2012ApJ...754...83Ba,2014ApJ...793..115Ba,2013MNRAS.432.3520D,2012ApJ...756..164Fa,2011MNRAS.417..717W}, the dust correction is expected to be negligible at redshifts $z >$ 8; the $z \sim$ 9 and 10 SFR function parameters are thus directly related to the uncorrected UV LFs at those redshifts.

\subsection{Method to derive the average $SFR-M_h$ relation}
Adopting the approach presented in \citet{2004MNRAS.353..189V}, we use the observed, dust-corrected $z \sim$ 4-8 luminosity functions, i.e. SFR functions, to derive an empirical relation between the galaxy SFR and its host halo mass. In this abundance matching method, the SFR-$M_h$ relation is calculated by setting the SFR of a galaxy hosted in a halo of mass $M_h$ to be such that the number of galaxies with a star formation rate greater than SFR equals the number of halos with mass greater than $M_h$ at a given epoch:

\begin{equation}
\int_{M_h}^\infty n(M'_h,z)dM'_h = \int_{SFR}^\infty \phi(\text{SFR'},z)d\text{SFR'},
\end{equation}
where the $\phi(\text{SFR},z)$ are the observed SFR function derived in the previous section and $n(M_h,z)$ is the Sheth-Tormen halo mass function \citep{1999MNRAS.308..119S},

\begin{equation}
n(M_h,z)dM=A\left(1+\frac{1}{\nu^{2q}}\right)\sqrt{\frac{2}{\pi}}\frac{\rho_m}{M}\frac{d\nu}{dM}e^{-\nu^2/2}
\end{equation}
with $\nu = \sqrt{a}\delta_c/[D(z)\sigma(M)]$, $a$ = 0.707, $A$ = 0.322 and $q$ = 0.3; 
$\sigma(M)$ is the variance on the mass scale $M$ (assuming the linear theory density power spectrum) 
while $D(z)$ is the growth favor and $\delta_c$ is the linear threshold for spherical collapse, which in a flat universe is $\delta_c$ = 1.686.

The abundance matching technique, as prescribed in equation (3), presupposes that the SFR is a monotonic function of the halo mass. This is supported by the observed trend that the clustering strength of galaxies at high redshift increases with their UV luminosity, similar to that of the clustering strength of halos increasing with mass \citep{2009ApJ...695..368L}. Hence, a scaling relation that assumes the SFR of a galaxy is a monotonically increasing function of the mass of its host halo is a reasonable starting point. Furthermore, our model assumes a one-to-one correspondence between galaxies and host halos. This neglects the substructure expected to exist within a halo, comprised mainly of halos that formed at earlier epochs and merged to become subhaloes of more massive hosts \citep{2002MNRAS.329..246B,2000ASPC..200...29W}. To include the contribution of this subhalo population and the galaxies they are potentially hosting, we modified the left-hand side of equation (3) to integrate over the $total$ mass function, i.e. the sum of the regular Sheth-Tormen halo mass function and the unevolved subhalo mass function taken from \citet{2004MNRAS.353..189V}. We found that because values for the subhalo mass function are generally lower than those of the halo mass function, the inclusion of this additional substructure has a negligible effect on the resulting scaling law; the subhaloes' contribution to the expression is most important at the low-mass end, and even there, its inclusion changes the results by $\lesssim$ 10\%. Therefore, for the sake of simplicity, we neglect multiple halo occupation and assume that each dark-matter halo hosts a single galaxy.  

 \begin{figure*}
\centerline{\includegraphics[width=350pt,height=300pt]{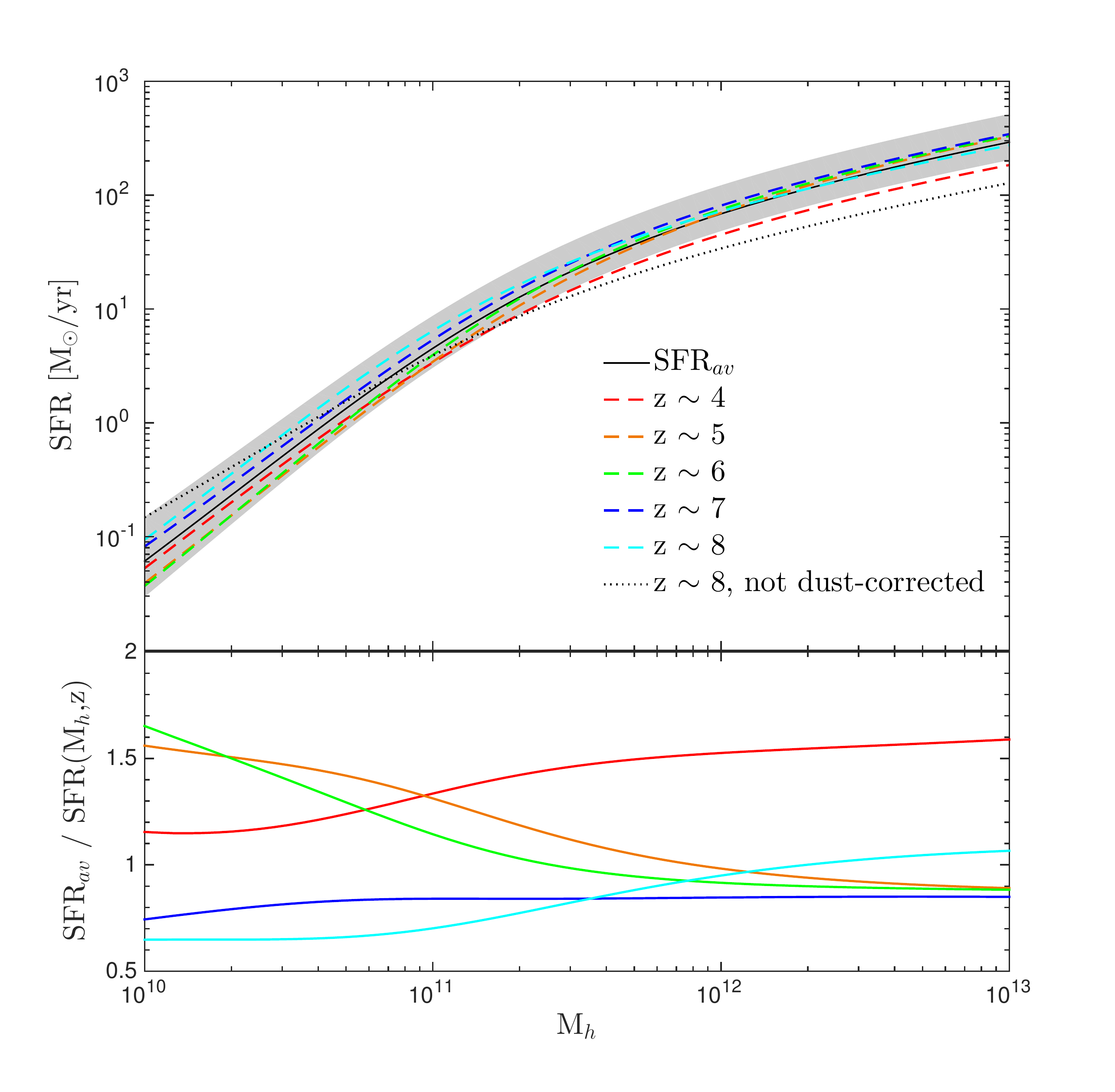}}
\caption{Top panel: the relation between galaxy SFR and dark matter halo mass, $SFR(M_h, z)$, derived at $z \sim$ 4-8 via abundance matching of the dust-corrected SFR functions to the halo mass function. For comparison, we show the $SFR-M$ relation at redshift $z$ = 8 obtained by abundance matching the \textit{observed} (uncorrected) UV LF function to the halo mass function at $z$ = 8 (dotted black line). The solid black curve represents the mean SFR-$M_h$ scaling law, $SFR_{av}(M_h)$, obtained by averaging the (dust-corrected) relations over the redshift range at a fixed halo mass. The uncertainty in our model calibration of $SFR_{av}(M_h)$ (shaded region), was derived by varying the $z \sim$ 4-8 SFR function parameters within their respective 1$\sigma$ confidence regions. Bottom panel: ratio of the mean scaling law to the SFR-$M_h$ relations derived at each redshift. The vertical axis represents the factor by which the average relation over/underestimates the $z \sim$ 4-8 scaling laws over the halo-mass range of interest.}
\label{fig:sfrplot}
\end{figure*}

\subsection{Modeling the SFR Functions}
Once the average relation between the SFR and the halo mass are known, we can model the SFR functions assuming a realistic dispersion around the average relation.
We model the probability density for a halo of mass $M_h$ to host a galaxy with a star formation rate $SFR$ to obey a log-normal distribution \citep{2001ApJ...550..177G,2003MNRAS.339.1057Y,2011ApJ...729...99M},
 \begin{multline}
P(\text{SFR}|M_h)d\,\text{SFR}=\\ \frac{1}{\sigma\sqrt{2\pi}}\exp{\left(-\frac{\log_{10}^2(\text{SFR}/\text{SFR$_{av}$}(M_h))}{2\sigma^2}\right)}d\log_{10}\text{SFR}
\end{multline}
where $SFR_{av}(M_h)$ is the mean SFR-$M_h$ relation derived by averaging the $SFR(M_h,z)$ relations obtained for 4 $\leq z \leq$ 8 over the specified range of redshifts at fixed halo mass. The variance, $\sigma^2$, represents the scatter in SFR at a fixed $M_h$ that arises from the stochastic nature of star formation activity. Halos of similar masses can have a range of large-scale environments, merger histories, and central concentrations \citep{2009ApJ...695..368L}. These conditions, along with interactions with nearby systems can result in different rates of gas accretion and star formation in galaxies. To account for this variance in the SFR-$M_h$ relation, we adopt a constant intrinsic scatter of 0.5 dex. This choice is motivated by the 0.5 dex scatter in stellar mass at constant SFR found by \citet{2011ApJ...735L..34G}, a result further confirmed by \citet{2014MNRAS.439.1326W} as providing the best statistical fit to observations of the SFR-$M_h$ relation.

The SFR function, $\phi(SFR)$, can then be obtained by integrating over the probability-weighted number densities of all halos that can achieve the star formation rate SFR within the allowed scatter:
\begin{equation}
\begin{split}
&\phi(\text{SFR}, z)=\int dM_h \,n(M_h, z)\, P(\text{SFR}|M_h)\\
&= \frac{1}{\sigma\sqrt{2\pi}}\frac{1}{SFR}\int dM_h\, n(M_h,z)\,e^{{-\frac{\log_{10}^2(\text{SFR}/\text{SFR$_{av}$}(M_h))}{2\sigma^2}}} \,.
\end{split}
\end{equation}
Note that in the limit of the variance $\sigma^2\rightarrow$ 0, the log-normal probability distribution becomes a delta function and the equation is reduced to the simpler form,
\begin{equation}
\phi(\text{SFR}, z)=n(M_h,z)\,\Bigg|\frac{d\,\text{SFR}_{av}(M_h)}{dM_h}\Bigg|^{-1} \,\,\, .
\end{equation}

 \begin{figure*}
\subfloat {\hspace{-.5 cm} \includegraphics[width=300pt]{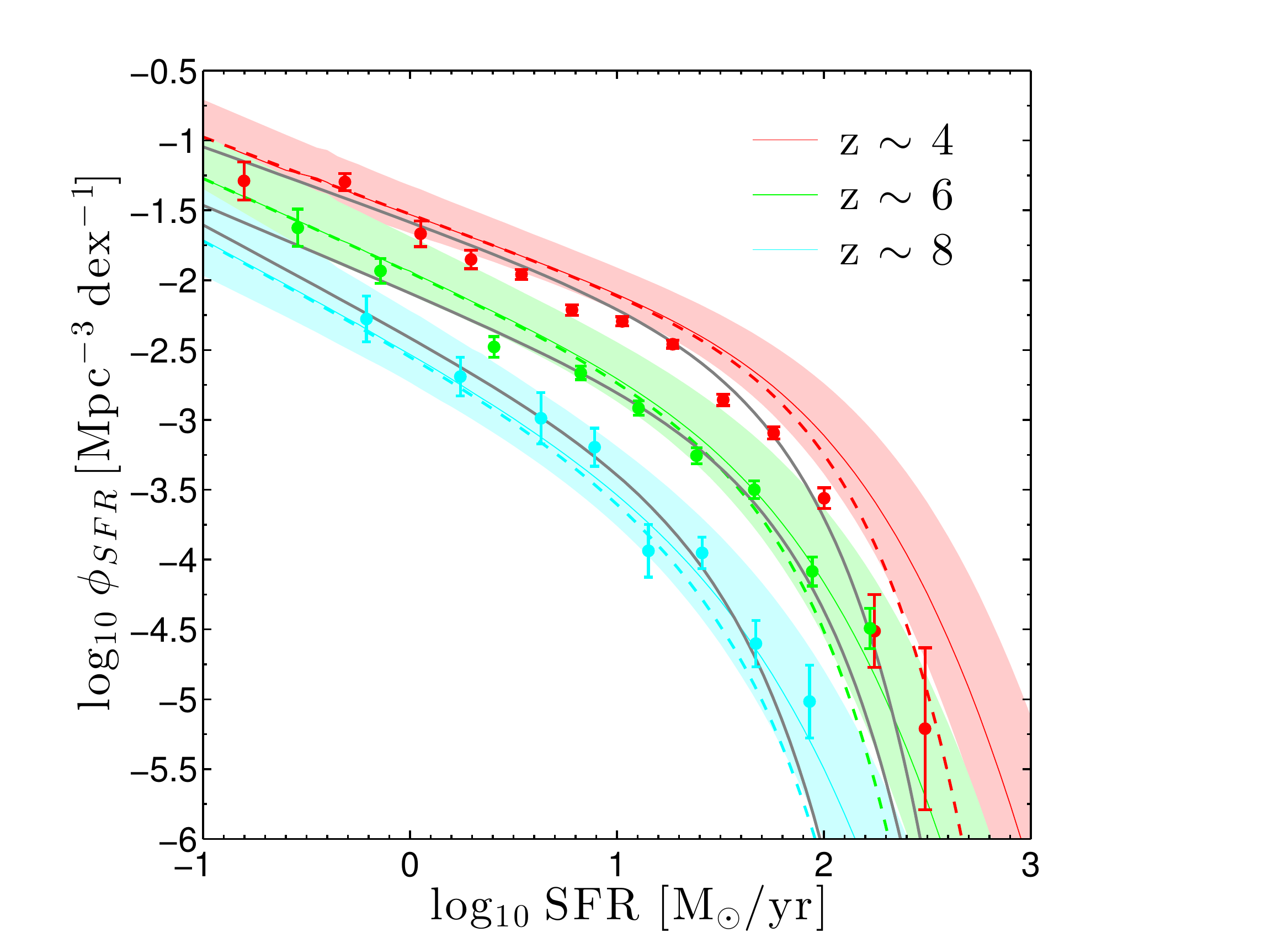}}
\subfloat{\hspace{-1 cm}\includegraphics[width=300pt]{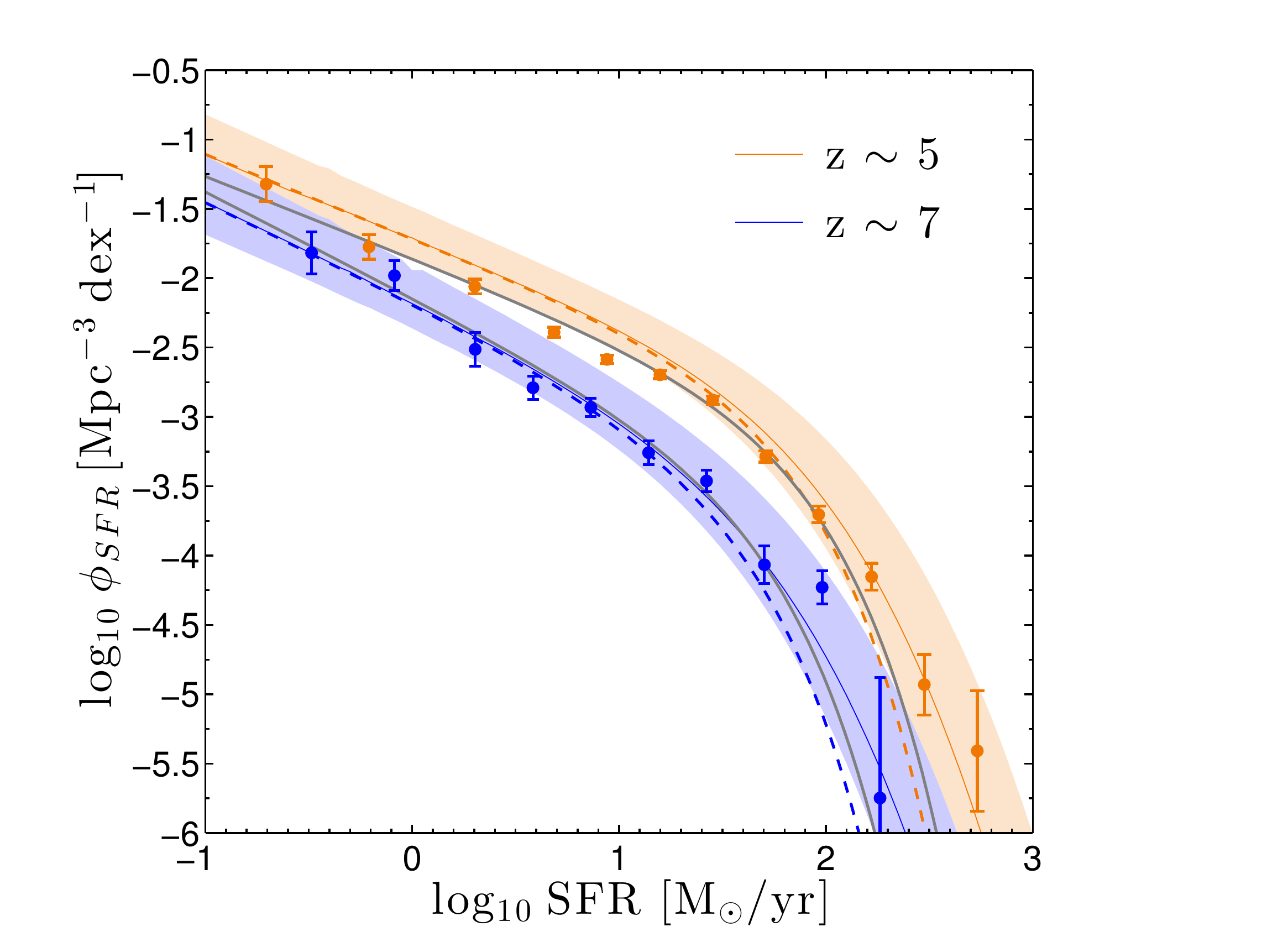}}\\
 \caption{Comparison of the predicted SFRFs at $z\sim4-8$ assuming an intrinsic scatter $\sigma$ = 0.5 dex (equation (6); solid colored curves) with those obtained by setting the variance equal to zero (equation (7); dashed curves). Our model SFRFs are overplotted with the data points at each respective redshift, along with the Schechter functions found to best approximate these measurements (gray curves; parameters recorded in Table 1). The data points represent the intrinsic UV luminosities, corrected for dust extinction and converted to SFRs using equation (1); the SFR functions are thus the intrinsic, dust-corrected LFs at these redshifts. The shaded areas in each panel correspond to the 1$\sigma$ confidence regions for our model LFs (when intrinsic scatter is accounted for).}
\label{fig:sfrf1}
 \end{figure*}

\section{Results}

\subsection{The Constant $SFR-M_h$ Relation}
The curves in the top panel of Figure \ref{fig:sfrplot} depict the SFR-$M_h$ relations inferred from the observed LF functions at redshifts 4 $\leq z \leq$ 8 via the abundance matching technique discussed in \S2.1. 
The scaling law remains roughly constant within 0.2 dex across this redshift range and is fairly well fit by a double power law with a turnover at a characteristic halo mass of $M_h^*$ $\approx$ 2$\times$10$^{11}$ M$_\odot$. Averaging the relations over $z \sim$ 4-8 at a fixed halo mass, we obtain a mean scaling law, $SFR_{av}(M_h)$ (solid black curve), characterized by a power-law slope $\beta \sim$ 0.9 at the high-mass end, i.e. $M_h \gtrsim$ $M_h^*$, and $\beta \sim$ 1.5 at the faint end where $M_h$ falls below the characteristic mass.

These results are consistent with the models presented in \citet{2010ApJ...714L.202T} where the $L-M_h$ relation is calibrated at $z \sim$ 6 and a $L \sim M_h^\beta$ relation is derived with $\beta \sim$ 1.3-1.6 at the low-mass end. This steepening of the SFR-$M_h$ relation towards lower masses may be due to the increased importance of feedback mechanisms in the star formation activity, such as supernovae and stellar winds \citep{2001ApJ...550..177G}. Meanwhile, the overall flattening of $SFR(M_h)$ at these redshifts compared to the local universe, where SFR $\propto M_h^4$ is mainly a consequence of the steepening faint-end slope of the observed LF at these redshifts, where $\alpha <$ -1.5 \citep{2005ApJ...627L..89C}.

The uncertainty in our model calibration of $SFR_{av}(M_h)$, represented by the shaded region in Figure \ref{fig:sfrplot}, is derived by varying the $z \sim$ 4-8 SFR function parameters within their respective 1$\sigma$ confidence regions and rederiving the average SFR-$M_h$ relation. As can be seen, the dashed curves defining the SFR-$M_h$ relation at each redshift  fall well within the boundaries of the uncertainty in the average scaling law, with the exception of the $z \sim$ 4 relation at the high-mass end. The bottom panel of Figure \ref{fig:sfrplot} illustrates the accuracy with which the mean scaling law approximates SFR as a function of halo mass at different redshifts. Overall, the $SFR_{av}(M_h)$ relation provides a reasonable representation of the $SFR(M_h,z)$ relations at all redshifts, with departures of less than 0.2 dex within the halo mass range of interest. With the exception of the $z \sim$ 4 relation, the $SFR(M_h,z)$ are all very similar (within 0.5 dex) at $M_h >$ 10$^{12}$ M$_\odot$.
At the low-mass end, the $z \sim$ 5-6 and $z \sim$ 7-8 scaling laws are over- and under-estimated respectively by less than a factor of two, while the SFR at $z \sim$ 4 is overestimated over the full halo-mass range by approximately the same factor. We therefore expect the mean relation to yield accurate estimates of the SFR function at these redshifts and, assuming that the SFR-$M_h$ relationship continues to remain roughly unchanged for $z >$ 8, at higher redshifts as well.

   \begin{figure*}
\subfloat {\hspace{-.5 cm} \includegraphics[width=300pt]{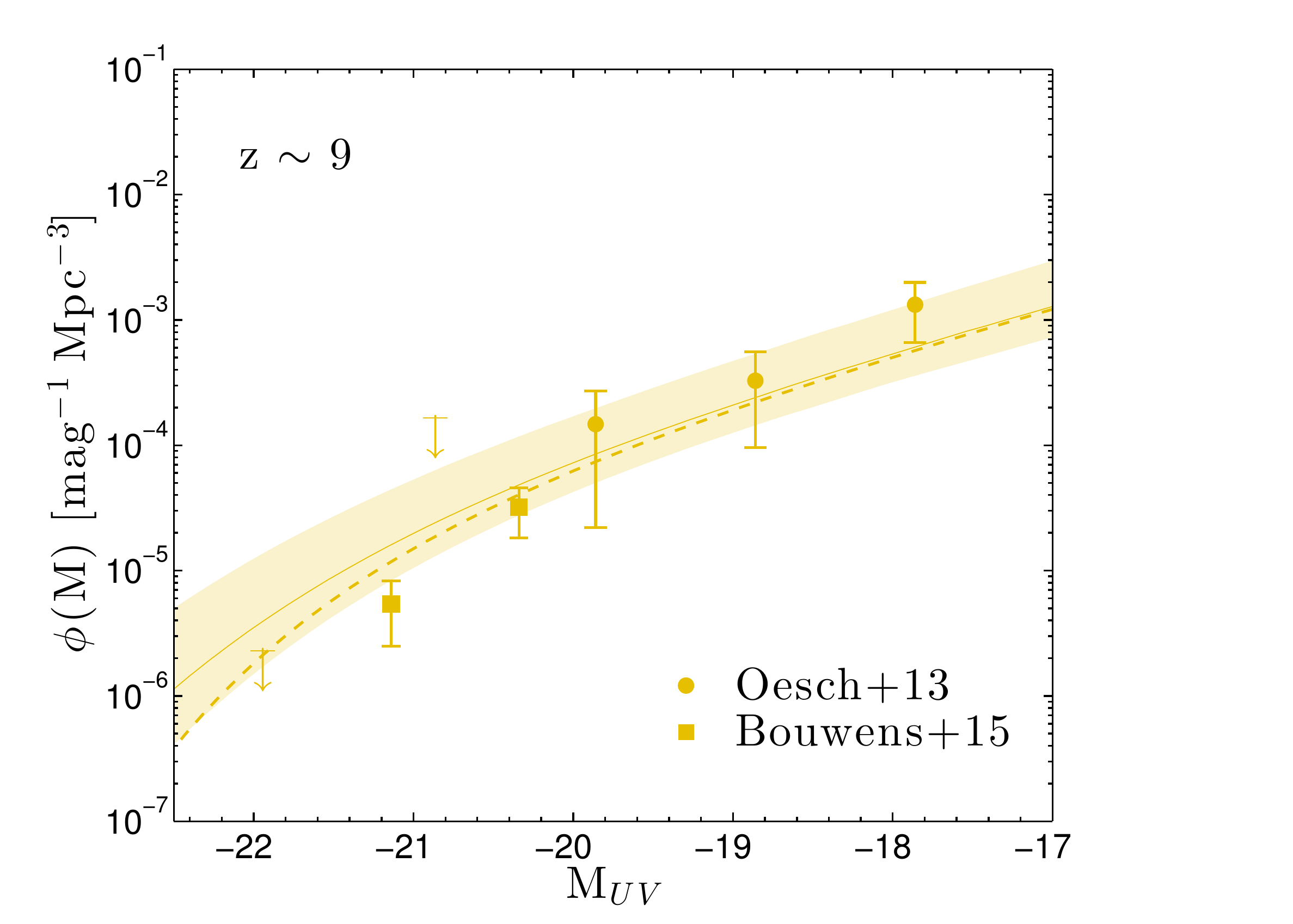}}
\subfloat{\hspace{-1 cm}\includegraphics[width=300pt]{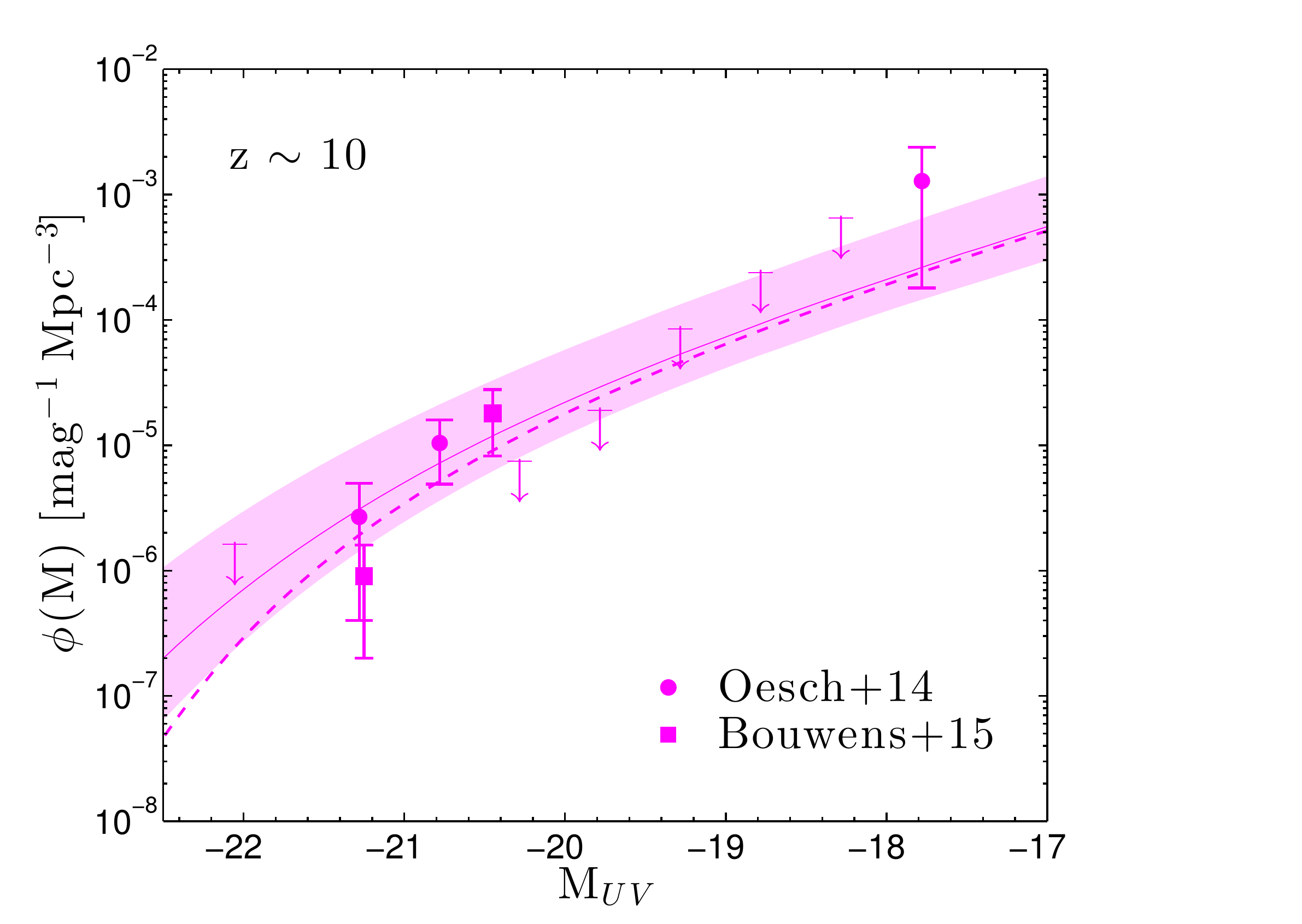}}\\
 \caption{Comparison of the predicted LFs assuming an intrinsic scatter $\sigma$ = 0.5 dex (equation 6; solid colored curves) with those obtained by setting the variance equal to zero (equation 7; dashed curves). Since at $z \sim$ 9 and 10, the dust extinction is expected to be negligible, the curves represent the observed, uncorrected LFs. We include determinations of the $z \sim$ 9 and $z \sim$ 10 UV LFs from \citet{2013ApJ...773...75O}, \citet{2014ApJ...786..108O}, and \citet{2015arXiv150601035B}. Arrows signify upper bounds on the number density at a given absolute UV magnitude, M$_{UV}$. The shaded area in each panel represents the 1$\sigma$ confidence region for the model LF when intrinsic scatter is accounted for. Our model UV LFs are in excellent agreement with the current observed estimates at these redshifts (albeit the latter are still based on small samples).}
\label{fig:sfrf2}
 \end{figure*}

\subsection{The Modeled SFR Functions at $z=4-8$}
Applying $SFR_{av}(M_h)$, we derive the expected SFR functions at redshifts $z \gtrsim$ 4 assuming only evolution of the underlying dark-matter halo mass function. The results of our LF model are shown in Figure \ref{fig:sfrf1}, both for the case where an intrinsic scatter of  $\sigma \sim$ 0.5 dex is assumed (solid colored curves) and in the limiting case where the variance is set to zero (dashed curves). Plotted alongside these results are the dust-corrected data points and the Schechter LFs with the observed best-fit parameters reported in Table 1 (gray curves). We find that including an intrinsic scatter increases the number density and yields SFR functions that are more consistent with observation, particularly at the luminous end where the increase in $\phi(SFR)$ due to scatter is most apparent. This boost in number density with the introduction of scatter can be attributed to the shape of the halo mass function. Even though $SFR(M_h)$ is log-normally distributed around the mean, $SFR_{av}(M_h)$, and the SFR scatter can thus go in either direction, the net change in the LF will always be dominated by low-mass halos entering into the galaxy sample by scattering into a SFR higher than its mean value, leading to a larger estimate of $\phi(SFR)$.

With the exception of $z \sim$ 4, our predicted SFR functions accurately reproduce the observed ones and provide very good fits to the data points at the luminous end when intrinsic scatter is accounted for. At $z\sim4$ our model predicts a SFRF which lies significantly above the observed data points by factors of 2-10$\times$. This is likely due to a combination of two effects: (\textit{i}) As also argued by \citet{2012ApJ...756...14S}, the bright end of the {\it observed} $z\sim4$ SFRF is likely underestimated due to heavily obscured galaxies, which are missed in UV-selected Lyman break galaxy samples; and (\textit{ii}) quiescent galaxies start to appear in the Universe at $z\sim4$ in significant numbers \citep[e.g.][]{Straatman14}. Since our model does not include any prescription to quench star formation in galaxies, this is likely another reason why the model appears to be inadequate at $z\sim4$. We conclude that our model should only be used at $z >$ 4, where it provides a very good representation of the observed SFRFs.


\subsection{A Prediction for the UV LF at $z=9-10$}

Building on the successful calibration of the SFR-$M_h$ relation at $z\sim5-8$, we use our model to predict the galaxy UV luminosity function at $z\sim9-10$. This represents the most distant galaxies in reach with HST, and recent observations have started to provide the first LF estimates at these redshifts. However, the sample sizes are still very small resulting in highly uncertain LF parameters \citep[see e.g.][]{2013ApJ...773...75O}. 

We derive the model UV LFs at $z$ = 9 and 10 by converting the SFR-$M_h$ relation calibrated in the previous section to a $L_{UV}-M_h$ relation using the standard \citet{1998ARA&A..36..189K} SFR-$L_{UV}$ relations. As can be seen in Figure \ref{fig:sfrf2}, our model provides an excellent fit to the current, albeit scarce, data. In particular, our model explains the somewhat puzzling absence of $z\sim10$ galaxy candidates between $M_{UV}=-18.5$ and $-20.5$ in current datasets. 
Incoming observations of the Hubble Frontier Field initiative will allow us to test this model further at $z\sim9-10$ in the future. 

 \begin{figure}
\vspace{-2cm}\includegraphics[width=84mm]{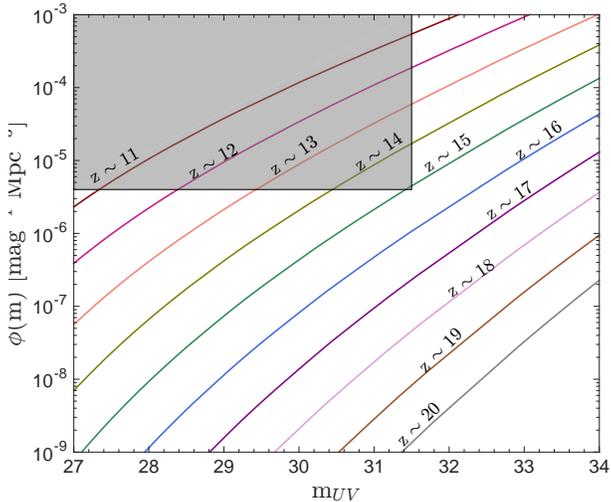}
\vspace{-2.5cm}\caption{Predicted LFs for $z >$ 10 based on our empirical model. The shaded region indicates the observable magnitude and volume within reach of a 200 arcmin$^2$ JWST treasury sized ultra-deep survey ($m_{lim} \simeq$ 31.5 mag). This indicates that the limiting redshift range observable with JWST may be $z\sim15$ for unlensed galaxies.}
\label{fig:sfrfhighz}
\end{figure}

\subsection{Extrapolation to $z\sim20$ and Predictions for JWST}
Motivated by the good agreement obtained by applying an unevolving SFR-$M_h$ relation to reproduce the observed LFs at 5 $\lesssim z \lesssim$ 10, we extend our approach to even higher redshifts for JWST predictions. The predicted LFs for redshifts $z >$ 10, again assuming a constant intrinsic scatter of $\sigma$ = 0.5 dex, are shown in Figure \ref{fig:sfrfhighz}. These model LFs, which are well-fitted by Schechter functions across the redshift range 4 $\lesssim z \lesssim$ 20, significantly evolve with redshift. The drop in number density from $z \sim$ 4 to 20 is reflected in the evolution of the characteristic number density, $\phi^*$, which decrease as $d\log{\phi^*}/dz \sim$ -0.3. We also find a gradual steepening of the faint-end slope ($d\alpha/dz \sim -0.08$) and a shifting of the characteristic luminosity towards fainter values for increasing redshift ($dM^*_{UV}/dz \sim$ 0.4). These trends in Schechter parameters are quite consistent with results from previous studies and extrapolations of observed $z \lesssim$ 10 LFs, except that our model $M^*$ evolves more significantly than found in observations \citep[e.g.,][]{2010ApJ...714L.202T,2011ApJ...737...90Ba, 2012ApJ...754...83Ba,2013MNRAS.432.2696M, 2014arXiv1410.5439F}. 

The shaded region in Figure \ref{fig:sfrfhighz} corresponds to the comoving volume and magnitude range accessible with a 200 arcmin$^2$ treasury sized ultra-deep JWST survey ($m_{lim} \simeq$ 31.5 mag). According to our model predictions, the limiting redshift for galaxy observations with a very deep JWST survey is $z\sim15$ \citep[see also][]{2006NewAR..50..113W}.
At higher redshifts, where the surface density drops below $\sim$10$^{-6}$ Mpc$^{-3}$, wider, deeper surveys would be required to observe the luminosity function of these rare, faint objects.

\begin{figure}
\vspace{-2.3cm}\hspace{-.5cm}\includegraphics[width=90mm]{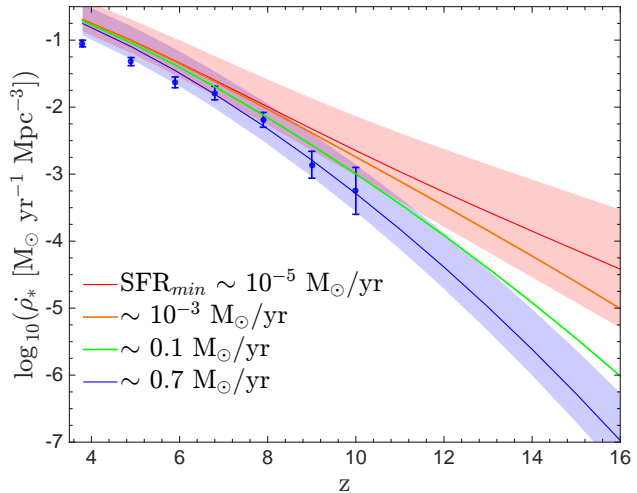}
\vspace{-3cm}\caption{The redshift evolution of the star formation rate density (SFRD) $\dot{\rho_*}$ derived by integrating the model SFR functions down to different star-formation limits, ranging from $SFR_{min} \sim$ 10$^{-5}$ to 0.7 M$_{\odot}$/yr. The shaded blue and red regions represents the 1$\sigma$ uncertainty in the SFRD obtained when integrating down to M$_{UV} \sim$ -17.7 mag (i.e., $SFR_{min} \sim$ 0.7 M$_{\odot}$/yr) and  M$_{UV} \sim$ -5.7 mag (i.e., $SFR_{min} \sim$ 10$^{-5}$ M$_{\odot}$/yr), respectively. Our model results (blue curve) can thus be compared with the (dust-corrected) measurements of the cosmic SFRD (blue circles) derived from this dataset. We find that the best-fit evolution of the SFRD at 8 $\leq z \leq$ 10 is $\dot{\rho_*} \propto$ (1+z)$^{-10.4\pm0.3}$, significantly steeper than the lower redshift trends which fall as (1+z)$^{-4.3}$.}
\label{fig:sfrd}
\end{figure}

Figure \ref{fig:sfrd} shows the model predictions for the redshift evolution of the SFR density, obtained by integrating the SFR functions down to different SFR limits, ranging from $SFR_{min} \sim$ 10$^{-5}$ M$_{\odot}$/yr to 0.7 M$_{\odot}$/yr. A minimum SFR of 0.7 M$_{\odot}$/yr corresponds to $M_{UV} \sim$ -17.7 mag, the magnitude of the faintest object observed in the HUDF12/XDF data; integrating down to this limit thus facilities comparison with the most recent, dust-corrected measurements of the SFRD, which have been plotted alongside the predicted curves in Figure \ref{fig:sfrd}. While our model appears to overpredict the SFRD measured for $z \lesssim$ 6 by $\sim$ 0.2-0.3 dex, the estimates of the $\dot{\rho_*}$ at higher redshifts agree quite well with observations.
 Furthermore, the evolution of the cosmic SFRD in our model is consistent with previous published results: the dust-corrected SFRD values evolve as $(1+z)^{-4.3\pm0.1}$ at 3 $< z <$ 8 before rapidly declining at higher redshifts where $\dot{\rho_*} \propto (1+z)^{-10.4\pm0.3}$ for 8 $\leq z \leq$ 10. In addition to accurately reproducing the order of magnitude drop in SFRD from $z <$ 8 to $z <$ 10 that has been observationally inferred in several separate analyses \citep{2013ApJ...773...75O}, our model finds that the SFRD continues to steeply decline towards higher redshift, predicting a cosmic SFRD of $\log{\dot{\rho_*}} \sim$ -7.0$\pm$0.3 M$_\odot$ yr$^{-1}$Mpc$^{-3}$ at $z \sim$ 16.

However, as will be shown and discussed in the following section, a star-formation rate density that declines as $(1+z)^{-10.4}$ at high redshifts, as derived assuming a minimum SFR of 0.7 M$_\odot$/yr, fails to reproduce the observed Planck optical depth.
Furthermore, galaxies are expected to exist beyond the current, observed magnitude limit. Theoretical and numerical investigations indicate that a halo at $z \leq$ 10 irradiated by a UV field comparable to the one required for reionization needs a minimum mass of $M_h \sim$ 6$\times$10$^7$ M$_{\odot}$ in order to cool and form stars \citep{1996ApJ...464..523H, 1997ApJ...474....1T}. 
We therefore explore the implications of this prediction by integrating the SFR density down to minimum SFRs as low as 10$^{-(5.0\pm0.8)}$ M$_\odot$\,yr$^{-1}$, the star formation rate corresponding to the minimum halo mass based on our average SFR-$M_h$ relation. 
This leads to a more moderate decline of the cosmic SFRD towards higher redshift. In this case, the best-fit evolution for 8 $\leq z \leq$ 10 is $\dot{\rho_*} \propto (1+z)^{-7.3\pm0.5}$, with an estimated SFRD of $\sim$ 4$\times$10$^{-5}$ M$_\odot$ yr$^{-1}$Mpc$^{-3}$ at $z\sim16$, a mere $\sim$ 250 million years after the Big Bang.

\begin{figure*}
\vspace{-1.5cm}
\includegraphics[width=0.49\linewidth]{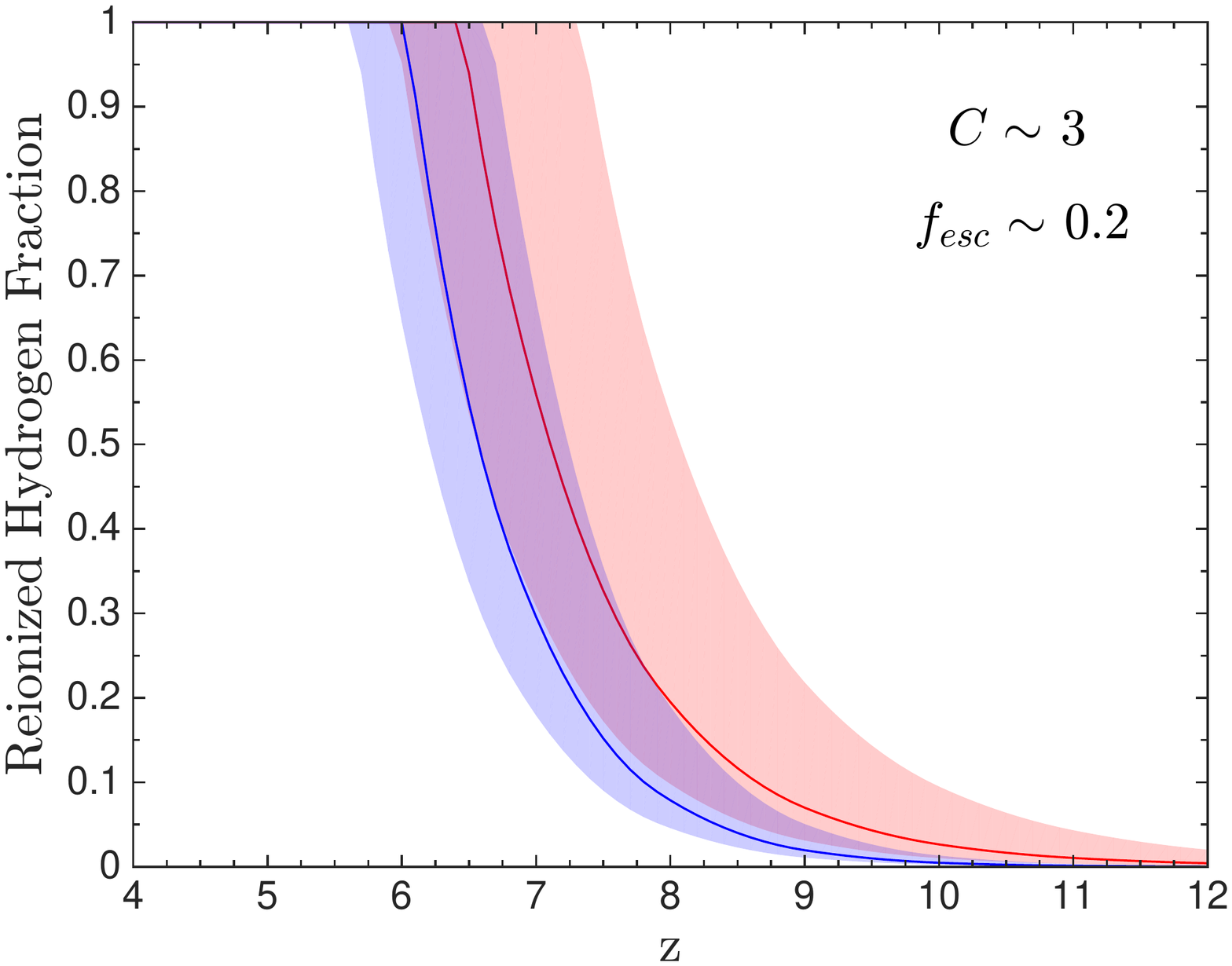}
\includegraphics[width=0.49\linewidth]{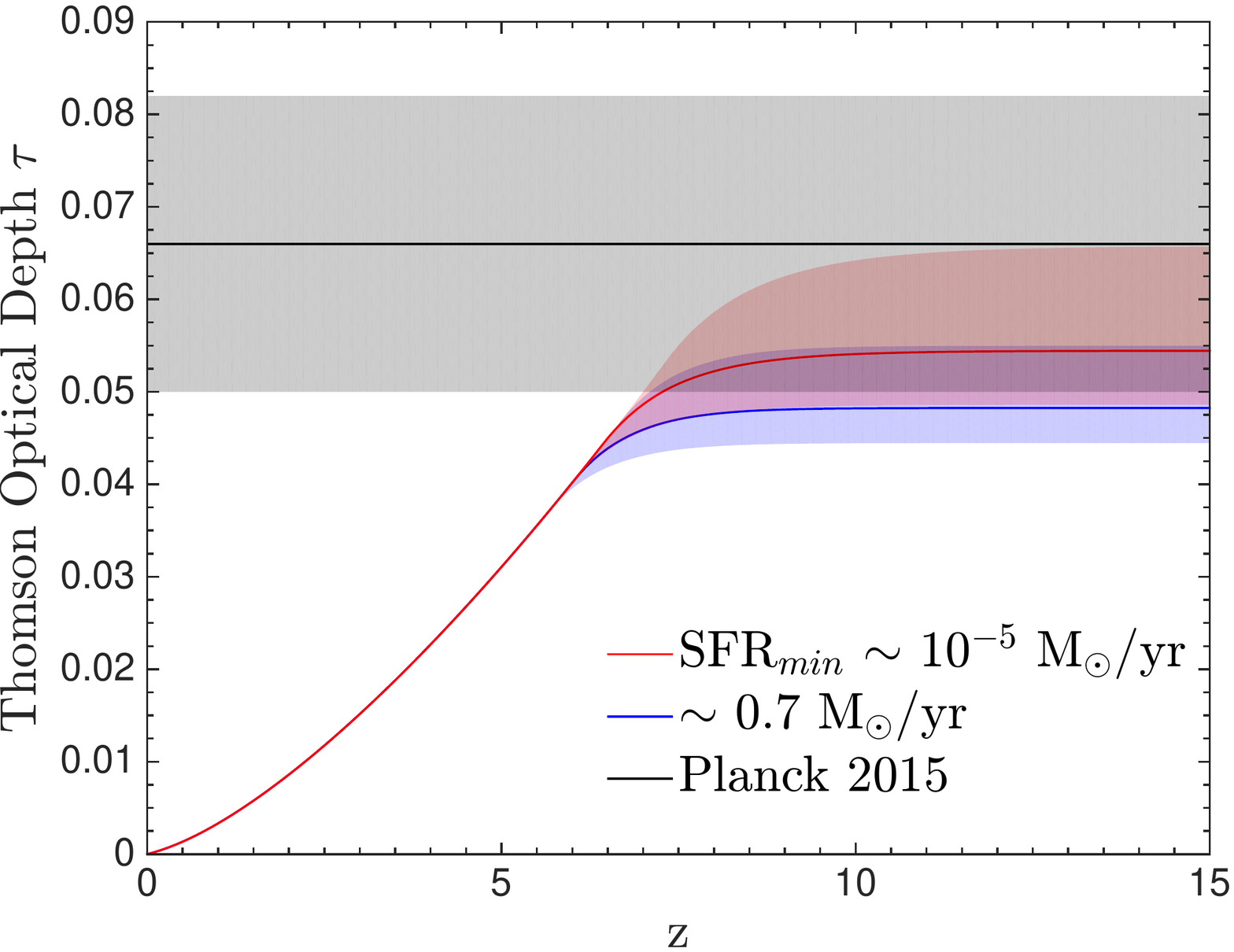}
\vspace{-2cm}\caption{Left panel: The reionization history $Q(z)$ calculated by solving the differential equation given by equation (8) assuming an ionizing photon escape fraction $f_{esc}$ = 0.2, an IGM clumping factor of $C$ = 3, and an average number of ionizing photons per unit mass of stars $N_\gamma$ = 7.65$\times$10$^{60}$ M$_\odot^{-1}$. The blue and red curves denote the reionization histories derived by using the cosmic SFRD $\dot{\rho_*}$ integrated down to $SFR_{min}$=  0.7 and 10$^{-(5.0\pm0.8)}$ M$_\odot$/yr respectively (i.e. M$_{UV} <$ -17.7 and $<$ -5.7 mag; galaxies hosted by the minimum halo mass to cool and form stars). For these chosen parameters, 68\% confidence limits on the redshift of half-reionization are 6.2 $< z <$ 7.3 when considering only the currently observed galaxy population, and 6.5 $< z <$ 8.1 with the inclusion of galaxies that expected to exist beyond this magnitude limit. Right panel: The corresponding Thomson electron scattering optical depths $\tau$ integrated over redshift from the present day, along with the Planck constraint $\tau$ = 0.066$\pm$0.016 (gray area). The shaded regions correspond to the 68\% confidence intervals, computed from the uncertainties in the cosmic SFRD $\dot{\rho_*}$ shown in Figure \ref{fig:sfrd}.}
\label{fig:reion}
\end{figure*}

\subsection{Contribution of Galaxies to Cosmic Reionization}
If star-forming galaxies supply the bulk of the photons that drive the cosmic reionization process, then our redshift-evolving SFRD $\dot{\rho_*(z)}$ can be used to determine the reionization history of the universe. The ionized hydrogen fraction $Q(z)$ can be expressed as a time-dependent differential equation \citep{loebbook},
\begin{equation}
\dot{Q}=\frac{\dot{n}_{ion}}{\langle n_H\rangle}-\frac{Q}{t_{rec}}
\end{equation}
where $\langle n_H \rangle$ is the comoving number density of hydrogen atoms and $\dot{n}_{ion}$ is the comoving production rate of ionizing photons, $\dot{n}_{ion}$ = $f_{esc}N_\gamma\dot{\rho_*(z)}$ where $N_\gamma$ is the number of ionizing photons per unit mass of stars and $f_{esc}$ is the average escape fraction from galaxies. The recombination time in the IGM is
\begin{equation}
t_{rec}=\left[C\alpha_B(T)(1+Y_p/4X_p)\langle n_H \rangle (1+z)^3\right]^{-1}
\end{equation}
where $\alpha_B(T)$ is the case B recombination coefficient for hydrogen (we assume an IGM temperature of 10,000 K corresponding to $\alpha_B \approx$ 2.79$\times$10$^{-79}$ Mpc$^3$yr$^{-1}$), $X_p$ = 0.75 and $Y_p$ = 1-$X_p$ are the primordial hydrogen and helium mass-fractions respectively, and $C$ is the clumping factor that accounts for the effects of IGM inhomogeneity. Estimates of the contribution of star-forming galaxies to reionization therefore depend on assumptions about the stellar IMF, metallicity, escape fraction, and clumping factor.

In this paper, we take $N_\gamma$ equal to 7.65$\times$10$^{60}$ M$_\odot^{-1}$, corresponding to a low-metallicity Chabrier IMF \citep{2015ApJ...799...32B}. The escape fraction is more difficult to constrain, especially at higher redshifts where the correction for intervening IGM absorption systems is high; while inferences at $z <$ 5 suggest $f_{esc} <$ 10\%, simulations suggest that $f_{esc}$ can be very large \citep{2009ApJ...693..984W, 2014MNRAS.442.2560W, 2011ApJ...730....8H} . For the sake of simplicity, we follow the approach in the previous literature and adopt a constant fiducial value of $f_{esc}$ = 0.2 for all redshifts. The determination of the clumping factor is also uncertain, with estimates ranging from $C$ = 2 to $C$ = 4 at high redshifts \citep{2009MNRAS.394.1812P, 2012MNRAS.427.2464F, 2012ApJ...747..100S}; we therefore assume $C$ = 3 in our fiducial model \citep[see also][]{2009ApJ...690.1350O,2012ApJ...752L...5Bb,2014arXiv1410.5439F, 2015ApJ...799...32B, 2015arXiv150308228B,2015ApJ...802L..19R}.

For a given reionization history $Q(z)$, the electron scattering optical depth $\tau$ can also be calculated as a function of redshift,
\begin{equation}
\tau(z)=\int_0^z c\langle n_H \rangle \sigma_T (1+Y_p/4X_p)Q(z')H(z')(1+z')^2dz'
\end{equation}
where $c$ is the speed of light, $\sigma_T$ is the Thomson cross section, and $H(z)$ is the Hubble parameter. This optical depth can be inferred from observations of the cosmic microwave background, and for a while, the accepted value from the Wilkinson Microwave Anisotropy Probe (WMAP) 9-year dataset was $\tau$ = 0.088$\pm$ 0.014, which, in the simplest model, corresponds to instantaneous reionization at $z_{reion} \simeq$ 10.5$\pm$1.1.
In early 2015, a significantly lower value of $\tau$ = 0.066$\pm$0.016 was reported \citep{2015arXiv150201589P}, consistent with instantaneous reionization occurring at $z_{reion} \simeq$ 8.8$^{+1.2}_{-1.1}$. More recently, the value has again shifted up to $\tau$ = 0.078$\pm$0.019, consistent with instantaneous reionization occurring at $z_{reion} \simeq$ 9.9$^{+1.8}_{-1.6}$ \citep{2015arXiv150702704P}.

Figure \ref{fig:reion} shows the expected reionization history of the universe (left panel) and the corresponding optical depth (right panel) computed using the cosmic SFRD $\dot{\rho_*}$ we derived\ from our model UV LFs. We find that the currently observed galaxy population at magnitudes brighter than $M_{UV} \leq$ -17.7 ($SFR \geq$ 0.7 M$_\odot$/yr, blue curve) fully reionize the universe by redshift $z \sim$ 6 and that the model corresponding to this magnitude limit predicts a Thomson optical depth of $\tau \simeq$ 0.048. Integrating further down to $M_{UV} \leq$ -5.7 ($SFR \geq$ 10$^{-5}$ M$_\odot$/yr, red curve) produces a model that reionizes the universe slightly earlier, at $z \sim$ 6.5, with a predicted optical depth of $\tau \simeq$ 0.054. The shaded regions corresponding to each of these curves represent the 68\% confidence limits arising solely from the uncertainties in the normalization of the cosmic SFRD.

Our results for the reionization history are consistent with measurements from high-redshift quasar and gamma-ray burst spectra, as well as from Lyman alpha emission in high-$z$ galaxies which collectively indicate that reionization ended near $z \simeq$ 6 \citep{2007MNRAS.382..325B,2012MNRAS.423..862K,2013ApJ...774...26C, 2013ApJ...775L..29T, 2014ApJ...795...20S, 2015MNRAS.447..499M}.
Furthermore, the model which includes the population of low-mass halos with SFRs as faint as 10$^{-5}$ M$_\odot$/yr produces an optical depth that is consistent with the previous inferred value by Planck, $\tau$ = 0.066 \citep{2015arXiv150201589P}, at the 1$\sigma$ level.
These findings, in line with previous analyses, indicate that a significant population of low-mass star-forming galaxies is necessary for cosmic reionization and strengthen the conclusion that the bulk of photons responsible for reionizing the early universe emerged from ultra-faint galaxies.

In order for our model to reproduce the most recent value of $\tau$ = 0.078 inferred from CMB observations by Planck, the average $SFR-M_h$ relation used would have to allow for larger star formation rates at the low-mass end. 
One simple way to satisfy such conditions is to permit $SFR_{av}(M_h)$ to vary according to the average relation derived via abundance matching for $M_h \gtrsim$ 10$^{11}$ M$_\odot$, but decline as a power law for $M_h \lesssim$ 10$^{11}$ M$_\odot$ with a slope of $\beta \sim$ 1.3, shallower than the slope derived empirically in \S3.1 at the low mass end. In such a model, the corresponding optical depth would be $\tau \sim$ 0.076, similar to the latest reduced measurement of $\tau$ = 0.078. While the high-redshift LF functions derived assuming such a form for the average $SFR-M_h$ relation are in tension with current observations, future measurements taken with JWST will shed more light on how the LF functions evolve at these low SFRs.

We also note that the results for $Q(z)$ and $\tau(z)$ depend on the choices for the escape fraction $f_{esc}$, the number of ionizing photons per unit stellar mass $N_\gamma$, and the clumping factor $C$; if one assumes a smaller clumping factor, or a higher value for $f_{esc}$ or $N_\gamma$, the evolution of $Q(z)$ will be shifted towards higher redshifts and an optical depth closer to the one measured by \citet{2015arXiv150702704P} will be obtained. Furthermore, the ionizing photons of Population III stars are expected to additionally increase the optical depth at high redshift \citep{2015arXiv150401734S}.

\section{Summary}

In this paper, we use the most recent $z \sim$ 4-8 UV LFs and UV continuum slope measurements to derive an empirical prediction of the evolution of UV LFs at $z >$ 8. 
Assuming a monotonic, one-to-one correspondence between the observed galaxy SFRs and the host halo masses, we map the shape of the observed $z \sim$ 4-8  UV LFs to that of the halo mass function at the respective redshifts. We find that the resulting $SFR-M_h$ scaling law remains roughly constant over this redshift range and is fairly well fit by a double power law, $SFR \propto M_h^\beta$, with $\beta \sim$ 0.9 at the high-mass end, i.e. $M_h \gtrsim$ 2$\times$10$^{11}$ M$_\odot$, and $\beta \sim$ 1.5 at the low-mass end. We note that the unevolving nature of this relation with redshift is an empirical result, an explanation for which lies beyond the scope of this paper. Future work based on semi-analytic models and numerical simulations, as well as anticipated observations with Atacama Large Millimeter Array (ALMA), are expected to shed further light on this matter. 

Applying this average $SFR-M_h$ relation with an intrinsic scatter of $\sigma \sim$ 0.5, we accurately reproduce the observed SFR functions at 5 $\lesssim z \lesssim$ 10 (Figure 3 and 4) and extend our approach to predict the evolution of the UV LF at redshifts $z \sim$ 11-20. We find an evolving characteristic number density $\phi^*$ which decreases as $d\log{\phi^*}/dz \sim$ -0.3, a gradually steepening faint-end slope, $d\alpha/dz \sim -0.08$, and a shifting of the characteristic luminosity towards fainter values, $dM^*_{UV}/dz \sim$ 0.4. Given the comoving volume and magnitude range of an ultra-deep JWST survey, our model predicts that observations of the LF up to $z \sim$ 15 are within reach in the absence of gravitational lensing, while deeper and wider surveys would be necessary to observe higher redshift objects. 

We also derive the evolution of the SFR density by integrating the SFRFs down to various SFR limits, ranging from 0.7 M$_\odot$/yr (corresponding to the faintest object observed in HUDF12/XDF), to 10$^{-5}$ M$_\odot$/yr (corresponding to the minimum halo mass necessary to cool and form stars). 
The inclusion of galaxies with SFRs well below the current detection limit results in a more moderately declining cosmic SFRD with redshift and leads to a fully reionized universe by $z \sim$ 6.5. 
Furthermore, the corresponding predicted optical depth in this model, $\tau \simeq$ 0.054, is consistent with the reduced value of $\tau$ = 0.066$\pm$0.016 inferred from CMB observations by Planck at the 1$\sigma$ level. 
These results strengthen the claim that a significant portion of the reionizing photons in the early universe were emitted by a population of low-mass, star-forming galaxies that have thus far evaded detection.

\section{Acknowledgements}
We thank R. Smit and R. Bouwens for helpful discussions regarding the SFRF measurements and for comments on an earlier version of this manuscript. This work was supported in part by NSF grant AST-1312034. This material is based upon work supported by the National Science Foundation Graduate Research Fellowship under Grant No. DGE1144152. Any opinion, findings, and conclusions or recommendations expressed in this material are those of the authors and do not necessarily reflect the views of the National Science Foundation.




\label{lastpage}

\end{document}